\documentclass[preprint]{revtex4-1}
\usepackage{bm}
\usepackage{amssymb}
\usepackage{scalerel}
\usepackage{epstopdf}
\usepackage{mathbbol}
\usepackage{amsmath}
\usepackage{graphics,graphicx,epsfig,ulem}
\usepackage{braket}
\usepackage{tikz}
\usepackage[titletoc]{appendix}

\begin{document} 

\title{Study of Rydberg blockade in thermal vapor}

\author{Arup Bhowmick}
\author{Dushmanta Kara}
\author{Ashok K. Mohapatra}
\email{a.mohapatra@niser.ac.in}
\affiliation{School of Physical Sciences, National Institute of Science Education and Research Bhubaneswar, HBNI, Jatni - 752050, India}

%\date{\today}%
\begin{abstract}
\textbf{%Rydberg blockade is normally observed in a cold atomic sample where the atoms are considered to be frozen to avoid decoherence during the excitation process. 
We present the experimental demonstration of Rydberg blockade in thermal atomic vapor where the atoms are not necessarily be frozen. We show that not all the interacting atoms but only the atoms with same velocity collectively participate in the blockade process. Using this observation, we formulated a suitable model based on super atom picture to study blockade interaction in thermal vapor. We performed an experiment to measure Rydberg population in rubidium thermal vapor using optical heterodyne detection technique and density dependent suppression of Rydberg population is observed in suitable experimental parameter regime. Further analysis of the experimental data using the model verifies the scaling law for van der Waals interaction strength $(C_{6})$ with principal quantum number of the Rydberg state with $11\%$ error. Our result suggests multi-photon excitation in thermal vapor with suitable laser configuration to probe Rydberg blockade interaction based optical nonlinearity and many body effects.} 
\end{abstract}
%This is the first ever quantitative verification of van der Walls blockade in thermal ensemble of atoms interacting with a cw laser. 
%\pacs{42.50.Nn, 32.80.Rm, 34.20.Cf, 42.50.Gy}
%\keywords{Atom-light interaction, Rydberg EIT, Optical non-linearity}%Use showkeys class option if keyword
                              %display desired
\maketitle
%\section{Introduction}
Cooperative phenomena mediated by strong Rydberg-Rydberg interaction provides useful application in quantum information processing using photons~\cite{saff10,jaks00,luki01}. It also can be used for realisation of single photon source~\cite{saff02,dudi12} and enhanced optical Kerr non-linearity which can be observed for single photons~\cite{peyr12,chan14,busc17}. Rydberg excitation in a dense frozen ensemble of atoms is able to shift the many-body Rydberg excited state outside the excitation line-width leading to a single excitation inside the blockade volume which has been studied extensively in ultra-cold atoms~\cite{tong04,sing04,heid07,rait08,urba09,gaet09,pohl10}. Coherent driving of the atoms in strong blockade regime leads the system to a many-body entangled state~\cite{dudi12r,webe15}. Theoretical study of the influence of the dissipation in blockade interaction induced many body effects have been reported~\cite{glae12,petr13,scho14}. Also, a recent theoretical study reveals the transition of the blockade phenomenon as a pure quantum system to a classical system due to the presence of dephasing~\cite{levi16}. An experimental study of strongly interacting cold Rydberg gas in dissipative regime has also been reported~\cite{malo14}.

Experiments with thermal vapor are attractive due to less complexity in the experimental set up in comparison with ultra-cold atoms. Recent experiments with thermal vapor shows a rich non-equilibrium phase transition in the mean field regime of the Rydberg interaction~\cite{carr13,siba16,lets17}. Van der Waal's interaction in thermal vapor has also been observed as a dephasing of the coherent Rabi oscillation of the thermal ensemble using nano-second pulse of the excitation laser~\cite{bale14}. Strongly correlated growth of the Rydberg aggregates due to interaction has also been observed~\cite{urvo15}. In these experiments, the thermal atoms were considered to be frozen during the excitation pulse. Partial suppression of Rydberg excitation as an evidence of Rydberg blockade has been reported in an experiment with thermal atomic beam~\cite{yosh17} where the excitation duration was much shorter than the Rabi oscillation period of the excitation laser. Hence, the multi-atom coherences due to formation of super-atom was absent in the system~\cite{yosh17}. 

In this article, we present the study of Rydberg excitation in thermal atomic vapor driven by cw-laser fields in strong interaction regime. Rydberg blockade was studied in steady state in the presence of decoherence due to thermal motion of the atoms and the multi-atom coherence due to super-atom formation. Based on a model with two interacting atoms in the Rydberg states, we show that the atoms moving with same velocity participate collectively in the blockade process. Whereas, the atoms in the blockade volume moving with different velocities behaves like non-interacting or may participate in the anti-blockade process as demonstrated recently~\cite{kara17}. A model based on the concept of super-atom in the presence of decoherence due to thermal motion for N-atoms in strong interaction regime is presented. Rydberg population in rubidium thermal vapor was measured using optical heterodyne detection technique (OHDT)~\cite{bhow16,bhow17} and suppression in Rydberg population in a suitable parameter regime is demonstrated. The analysis of the experimental data using the model is shown to be consistent with the scaling law of van der Waals interaction strength with principal quantum number of the Rydberg states.

\begin{figure}[t]
\includegraphics[angle=0,scale=0.3]{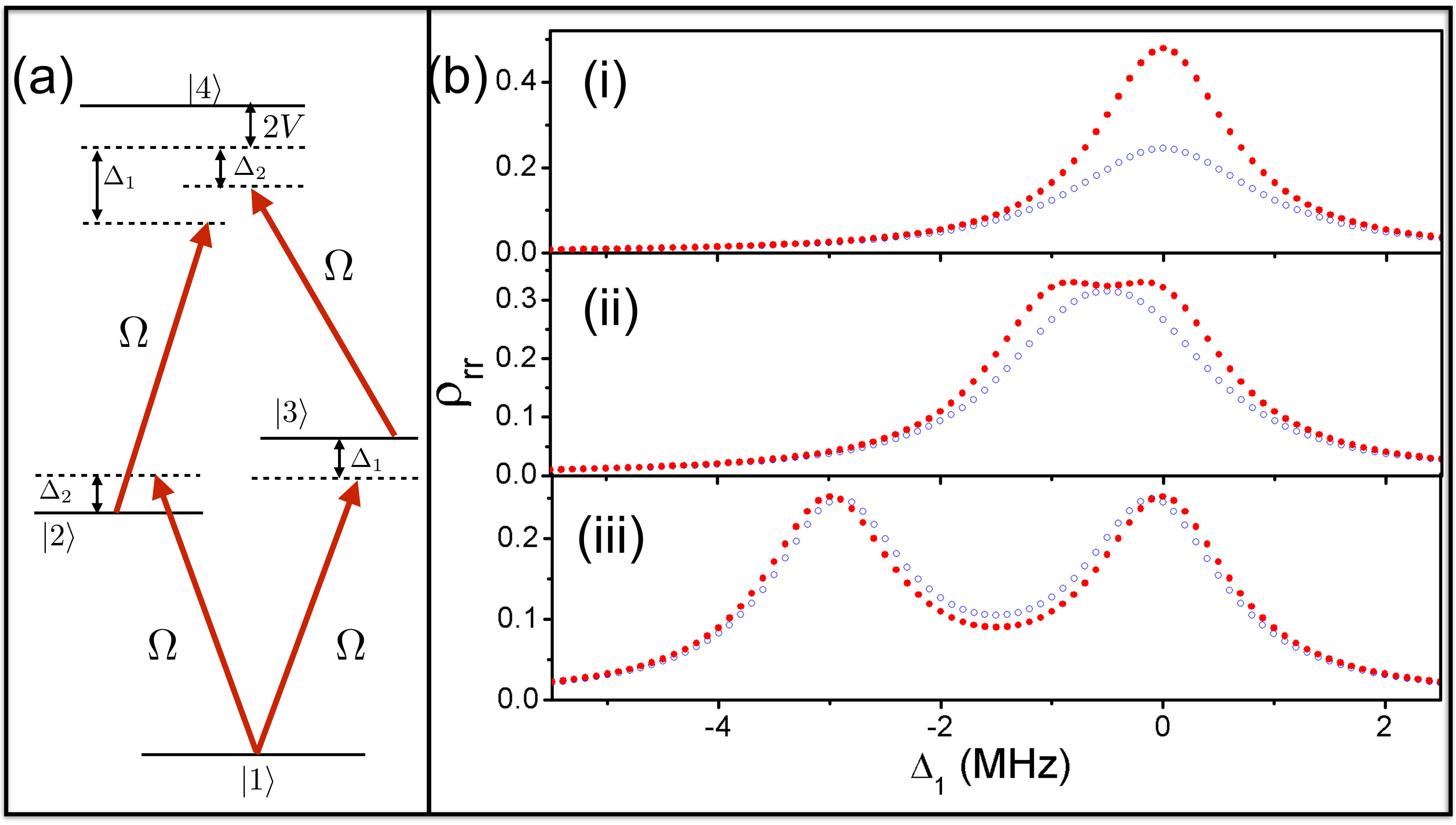}
\caption[]{(a) Energy level diagram of the composite system of two atoms. The state $\ket{1}$ corresponds to both the atoms in the ground state. The states $\ket{2}$ and $\ket{3}$ correspond to one atom in the ground state and the other atom in the Rydberg state. The state $\ket{4}$ correspond to both the atoms in the Rydberg state. (b) Rydberg population as a function of $\Delta_1$ (detuning from the first atom) calculated for two non-interacting atoms $({\bullet})$ and the atoms interacting in strong blockade regime $({\circ})$ with (i) $\Delta_1-\Delta_2=0$, (ii)   $\Delta_1-\Delta_2=\Omega$ and (iii) $\Delta_1-\Delta_2=3\Omega$. $\Omega$ used in the calculation to be $1$ MHz.}
\label{fig1}
\end{figure}

Schematic of the energy level diagram of the composite system of two atoms interacting in the Rydberg state is depicted in figure~\ref{fig1}(a). The Hamiltonian of the composite system is given by, $H=H^{(1)} \otimes \mathbb{1} +\mathbb{1} \otimes H^{(2)}+2V\ket{4}\bra{4}$ where $H^{(1)}$ and $H^{(2)}$ are the Hamiltonians of the individual atom interacting with the driving laser field with Rabi frequency $\Omega$ and detunings $\Delta_1$ and $\Delta_2$ respectively. Van der Waals interaction between two Rydberg atoms $\left(V=C_6/r^6\right)$ can be introduced as an energy shift of the state $\ket{4}$. The master equation of the combined system can be evaluated as, $\dot{\rho}=-\frac{i}{\hbar}[H,\rho]+\mathcal{L}_{D}(\rho)$ where $\rho$ is the density matrix of the composite system. The Lindblad operator accounting for the decoherence in the composite system can be written as $\mathcal{L}_{D}(\rho)=\mathcal{L}_{D1}\otimes \rho^{(2)} +\rho^{(1)} \otimes\mathcal{L}_{D2}$~\cite{bcgu13}. Here $\mathcal{L}_{Di}$ and $\rho^{(i)}$ are the Lindblad operator and density matrix for $i$th atom respectively. The only decoherence rate introduced in the problem is $\Gamma_{rg}$ accounting for the population decay from the Rydberg state to the ground state due to finite transit time of the thermal atoms. The master equation for the composite system was solved in steady state and Rydberg population was evaluated as $\rho_{rr}=\rho_{44}+\left(\rho_{22}+\rho_{33}\right)/2$. As shown in figure 1(b), if $\Delta_1=\Delta_2$, Rydberg population is strongly suppressed for the interacting atoms compared to the non-interacting atoms. Whereas, in the case of $\Delta_1-\Delta_2=\Omega$, Rydberg population of the interacting atoms doesn't change appreciably from the case of non-interacting atoms and becomes same for $\Delta_1-\Delta_2\gg\Omega$. Excitation to the Rydberg state in atomic vapor is usually performed with two laser fields with wave vector mismatch of $\Delta k$ depending on their relative alignment and the magnitude of the wave vectors. Collective excitation using blockade interaction is then possible for atoms with a difference in their velocities less than $\Omega/\Delta k$. In a thermal ensemble, the atoms within the blockade sphere resonating to the driving laser with a velocity width of $\Omega/\Delta k$ only collectively participate in the blockade interaction. The other atoms in the blockade sphere moving with different velocities either behave like non-interacting atoms or may contribute to the anti-blockade process.~\cite{kara17}. 

To model the blockade interaction in thermal vapor, we consider that the atoms with same velocity in a blockade sphere are collectively excited to the Rydberg state. An empirical formula based on super-atom model~\cite{dudi12r,webe15,heid07} has been deduced by considering the laser field interacting with a two-level atom with collective Rabi frequency as $\sqrt{N_b}\Omega$ with $N_b$ being the number of atoms inside the blockade sphere. To account for a single excitation inside the blockade sphere, the Rydberg population is divided with $N_b$ to find the effective population which is derived as, 
\begin{eqnarray}
\label{emp1}
\rho_{b}\left(N_b\right)=\frac{\Omega^2}{4\Delta^2+2N_b\Omega^2+\Gamma^2_{rg}}
\end{eqnarray}
The validity of the super-atom model is verified by comparing with the exact calculation for two and three atoms inside the blockade sphere as described in appendix~\ref{2am} and appendix~\ref{3am} respectively. The calculation was further extrapolated to N-interacting atoms inside the blockade sphere which is described in appendix~\ref{nam}. The empirical formula works well if the interaction shifts is assumed to be larger than the collective Rabi frequency.   

\begin{figure}[t]
\includegraphics[angle=0,scale=0.3]{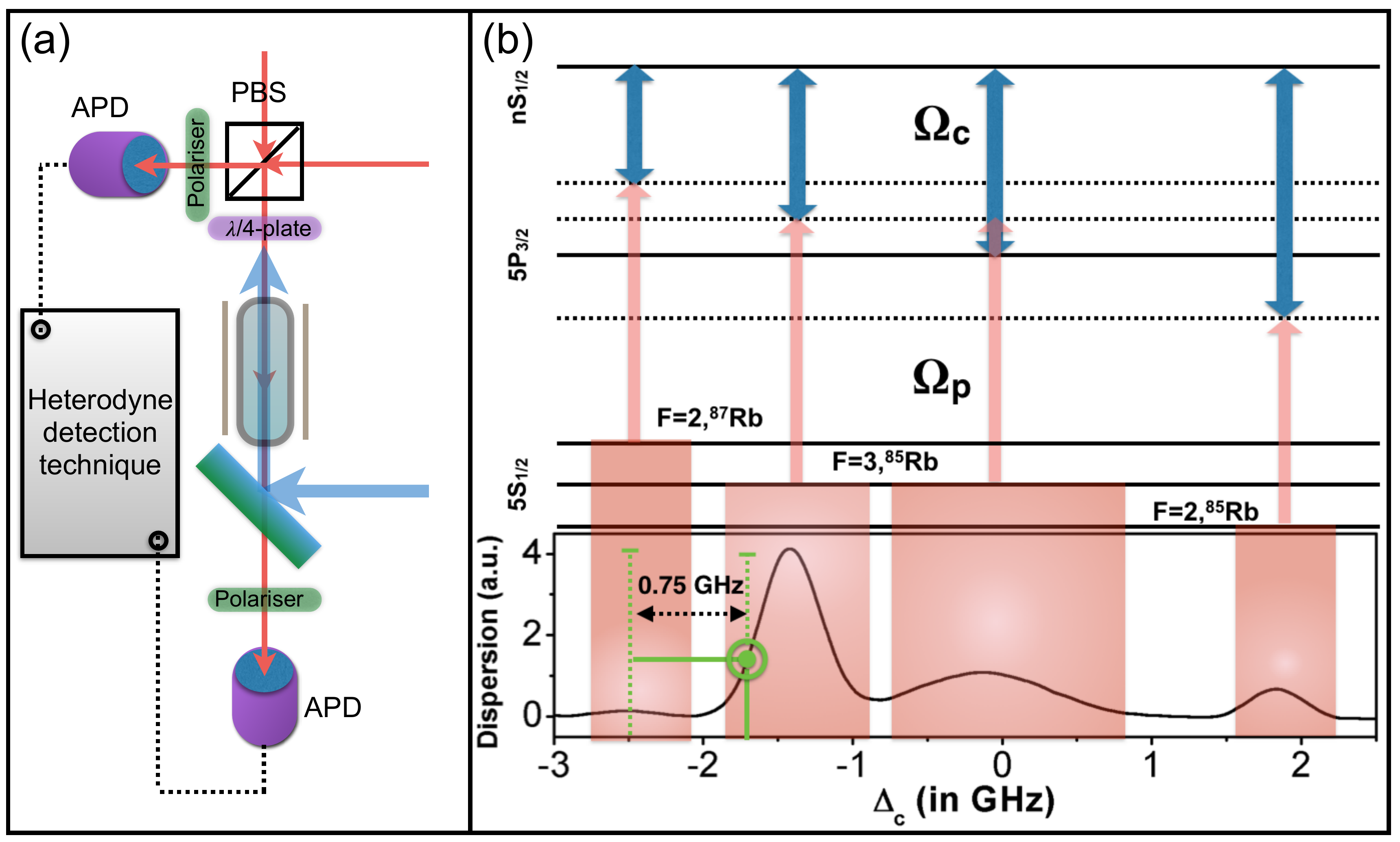}
\caption[]{(a) Schematic diagram of the experimental set up (b) Typical dispersion spectrum observed in the experiment by scanning the coupling laser over $5$ GHz. The resonance peaks corresponding to the two-photon resonances of $^{85}$Rb and $^{87}$Rb are depicted. The anti-blockade peak near $\Delta_c=0$ doesn't satisfy the usual two-photon resonance. Dispersion height at $0.75$ GHz blue detuned from the transition $5$s$_{1/2} (F=2)\rightarrow n$s$_{1/2}$ of $^{87}$Rb is used for further analysis to study the blockade interaction in thermal vapor.}
\label{fig2}
\end{figure}

\begin{figure}[t]
\includegraphics[angle=0,scale=0.3]{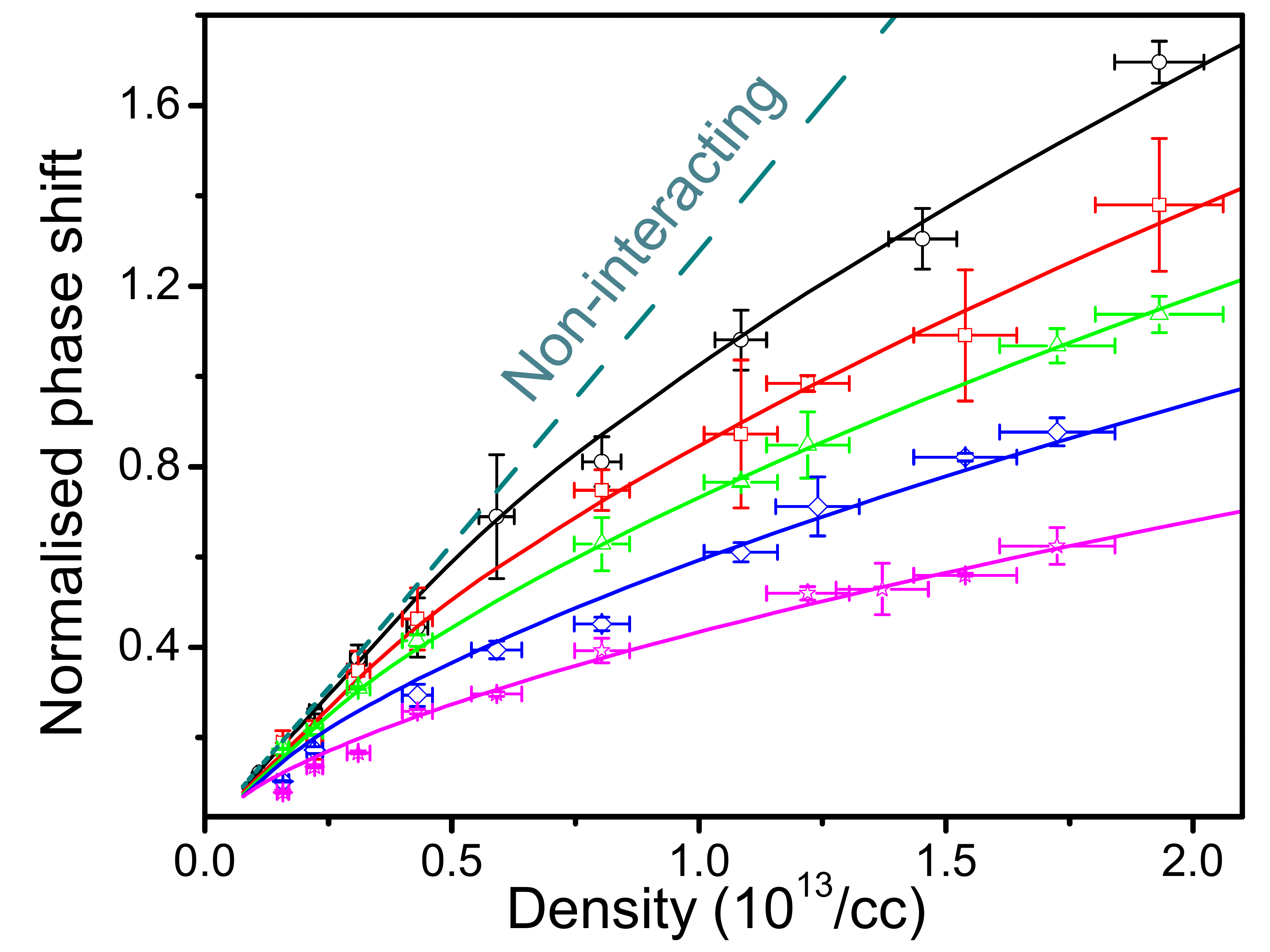}
\caption[]{Non-linear phase shift of the probe measured in the experiment due to two-photon transition to the Rydberg states with principal quantum numbers $n=35$ $\left({\scaleobj{1.6}{\circ}}\right)$, $n=40$ $\left({\scaleobj{0.8}{\square}}\right)$, $n=45$ $\left({\scaleobj{0.8}{\triangle}}\right)$, $n=50$ $\left({\scaleobj{1.6}{\diamond}}\right)$ and $n=53$ $\left({\scaleobj{1.6}{\star}}\right)$. The solid lines are the $\chi^{2}$-fitted curves using the model described in the text. The phase shift is normalised by a proportionality constant which accounts for the overall gain of the experimental set up. The dotted line is the expected signal for the non-interacting 
atomic ensemble.}
\label{fig3}
\end{figure}

Demonstration of blockade interaction was performed in our vapor cell experiment and the details of the experimental set up has been presented in our previous work~\cite{bhow16}. The schematic of the setup is depicted in figure~\ref{fig2}(a). In the experiment, a probe beam along with a reference laser beams were derived from an external cavity diode laser operating at wavelength of 780 nm. The frequency of the probe laser was locked at $1.2$ GHz blue detuned to the $5\text{S}_{1/2}\left(\text{F}=3\right) \rightarrow 5\text{P}_{3/2}$ transition of $^{85}$Rb. The coupling laser operating at 478 - 482 nm counter-propagates the probe laser through a magnetically shielded rubidium vapor cell. The cell was kept inside a oven and the temperature was controlled using a PID-controller to control the density of the vapor. The probe and coupling Rabi frequencies are denoted by $\Omega_p$ and $\Omega_c$ respectively which were determined from their respective intensities using the method reported in reference~\cite{bhow16}. $\Omega_p$ was kept constant by keeping the probe power fixed. The coupling Rabi frequency scales with principal quantum number as $\Omega_c\propto n^{3/2}$. The coupling laser power was adjusted accordingly to keep the coupling Rabi frequency constant throughout the experiment. Rydberg electromagnetically induced transparency signal~\cite{moha07} was optimized to ensure the overlapping between the probe and coupling beams. The beam waist of the probe (coupling) beam was $95$ $\mu$m ($80$ $\mu$m) and the respective Rayleigh range was $36.33$ mm ($41.86$ mm). The peak Rabi frequencies of the probe and coupling beams were  $400$ MHz and $8$ MHz respectively. The averaging due to the Gaussian intensity  profile of the beams were done in the theoretical model to compare with the experimental result. The probe and coupling detunings are denoted as $\Delta_p$ and $\Delta_c$ respectively.  The phase shift experienced by the probe due to two-photon excitation to the Rydberg state was measured using OHDT. The detail of the OHDT and the theoretical model to find relation between Rydberg population and measured dispersion can be found in references~\cite{bhow16, bhow17}. The dispersion peaks corresponding to the usual two-photon resonance $5\text{S}_{1/2}\left(\text{F}=3\right) \rightarrow n\text{S}_{1/2}$ and $5\text{S}_{1/2}\left(\text{F}=2\right) \rightarrow n\text{S}_{1/2}$ of $^{85}$Rb and $5\text{S}_{1/2}\left(\text{F}=2\right) \rightarrow n\text{S}_{1/2}$ of $^{87}$Rb are depicted in figure~\ref{fig2}(b). The dispersion peak corresponding to the transition $5\text{S}_{1/2} \left(\text{F}=3\right)\rightarrow n\text{S}_{1/2}$ of $^{85}$Rb is used for further analysis for demonstration of blockade interaction. Dispersion peak observed near $\Delta_c=0$ is due to enhanced Rydberg excitation induced by interaction and is called as anti-blockade peak~\cite{kara17}. The repulsive interaction of the Rydberg atoms in the $n$S$_{1/2}$ state leads to the anti-blockade peak which appears on the blue detuned side of the two-photon resonant peak as demonstrated recently~\cite{kara17}. The two-photon resonant spectrum contains the contribution from the blockade interaction due to the atoms resonantly interacting with the driving laser and also the anti-blockade due to the interactions of atoms with different velocities. However, for repulsive Rydberg-Rydberg interaction, anti-blockade dominates on the blue detuned side and has negligible contribution on the red detuned side of the spectrum~\cite{kara17}. To demonstrate the blockade interaction in our experiment, the dispersion was measured at $0.75$ GHz blue detuned to the peak of the  $5\text{S}_{1/2}\left(\text{F}=2\right)\rightarrow n\text{S}_{1/2}$ transition of $^{87}$Rb as shown in figure~\ref{fig2}(b). Since this point is on the red detuned side of the $5\text{S}_{1/2} \left(\text{F}=3\right)\rightarrow n\text{S}_{1/2}$ transition of $^{85}$Rb, the anti-blockade is expected to be negligible and hence only the blockade interaction is expected to dictate any non-linear dependence of density for the dispersion measurement. The dispersion at the highlighted point in figure~\ref{fig2}(b) was measured by changing the density of the atomic vapor. The measurement was repeated for the principal quantum numbers ($n=35$, $40$, $45$, $50$ and $53$) of the Rydberg excited states. A density dependent suppression of dispersion has been observed which becomes stronger with increase in principal quantum number as shown in figure~\ref{fig3}.

\begin{figure}[t]
\includegraphics[angle=0,scale=0.34]{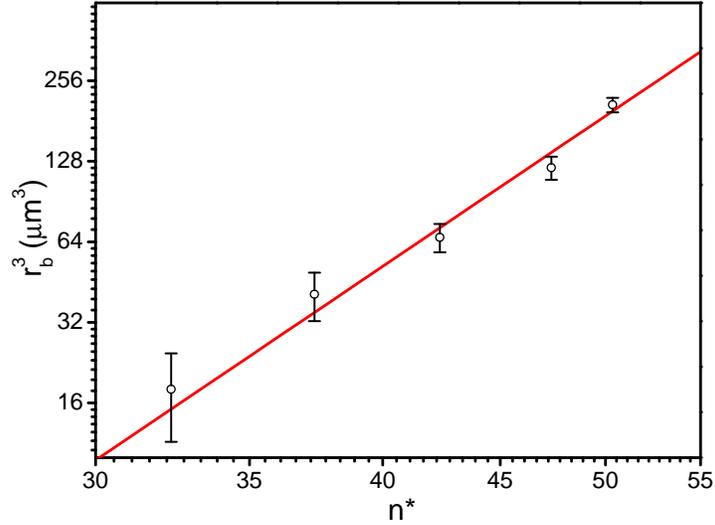}
\caption[]{$r^{3}_{b}$ as a function of $n^*$ where $n^*=n-\delta$ with $\delta$ being the quantum defect. The solid line is the fitting to find the scaling, $r^3_b\sim n^{* 5.8\pm 0.666}$.}
\label{fig4}
\end{figure}

The signal measured in the experiment using OHDT is proportional to the non-linear phase shift of the probe as $V_s\propto\Re\left(\chi_{3L}\right)$ where $\Re\left(\chi_{3L}\right)$ is the real part of the probe susceptibility due to two-photon excitation to the Rydberg state~\cite{bhow16}. Using the relation between $\Re\left(\chi_{3L}\right)$ with Rydberg population from reference~\cite{bhow17}, the signal measured in the experiment can be written as
\begin{eqnarray}
\label{fit_f}
V_s=n_0G\int^{\infty}_{-\infty} \xi(v) f(v)\rho_{eff} dv
\end{eqnarray}
where $n_0$ is the density of the medium. $G=\frac{2\mu_{eg}^2}{\epsilon_0 h}A_g$ where $A_g$ is the overall gain of the experimental set up. The Maxwell-Boltzmann velocity distribution is used as $f(v)=\frac{1}{\sqrt{\pi}v_p} e^{-v^2/v^2_p}$. In the regime, $\Omega_p\gg\Omega_c$ as used in the experiment, $\xi(v)\approx\frac{1}{2\left(\Delta_p-k_pv\right)}$ with $k_p$ being the probe wave vector. $\rho_{eff}$ is the effective Rydberg population which is evaluated using the expression~(\ref{emp1}) and following the method as explained in the Methods section. The effective detuning and Rabi frequency due to two-photon excitation are used in the expression~(\ref{emp1}) as $\Delta(v)=\Delta_p+\Delta_c+\Delta kv$ and $\Omega\approx\frac{\Omega_p\Omega_c}{2\Delta_p}$ respectively. The wave vecotr mismatch $\Delta k=k_c-k_p$ with $k_c$ being the coupling wave vector with counter-propagating configuration of probe and coupling beams. The number of atoms per blockade has been evaluated as, $N_{b} =\frac{4\pi}{3}r^3_bn_0 f(\tilde{v})\left(\frac{\Omega}{\Delta k}\right)$ with $r_b$ being the blockade radius, $\tilde{v}$ is the velocity of the atoms resonating with the lasers and $\frac{\Omega}{\Delta k}$ is the width in the velocity within which the atoms have significant contribution to the blockade interaction. Presence of the density term in $N_b$ has a significant role in evaluating the effective Rydberg population which is responsible for the observed density dependent suppression in the dispersion measurement. A quantitative comparison of the experimental data for the dispersion with the theoretical model was performed as explained below. The experimental parameters were kept constant throughout the experiment for all the principal quantum number states. So the proportionality constant $G$ was determined from the linear fitting of the low density data for $n=35$ where $N_b < 1$. Using the same fitted value of $G$, the expression~(\ref{fit_f}) is used to fit the data for all the principal quantum number states by using $r_b^3$ as the fitting parameter. The values determined for $r_b^3$ from the fitting is plotted as a function of $n^*$ which is shown in figure~\ref{fig4}. $r_b^3$ as a function of $n^*$ is further fitted to find the scaling as $r^3_b\sim n^{* 5.8\pm 0.666}$. Since the van der Waals interaction strength, $C_6\propto r_b^6$, then the scaling found in the experiment for the blockade radius is consistent with the scaling, $C_{6}\sim n^{*11}$ within $11\%$ error.
 
In summary, we have demonstrated Rydberg blockade in thermal atomic vapor. Collective behaviour of the atoms moving with same velocity inside the blockade sphere is observed while excited to the Rydberg state due to blockade interaction. The atoms moving with different velocity can behave non-interacting or contribute to the anti-blockade process under suitable laser parameter. To explain the spectrum due to Rydberg excitation in thermal vapor requires the model for the blockade interaction as well as anti-blockade process. The blockade radius can not be simply scaled by the Doppler width of the two-photon transition. The wave vector mismatch of the lasers used for two-photon excitation to Rydberg state in alkali vapor leads to the Doppler width of the order of few $100$ MHz which reduces the optical non-linearity due to blockade interaction. Multi-photon (three or four photons) transition to the Rydberg state allows to reduce the wave vector mismatch  by suitable alignment of the multiple lasers as demonstrated recently in reference~\cite{carr13}. In this case, all the atoms moving with different velocities can contribute to the collective excitation due to blockade interaction. Large optical non-linearity induced by blockade interaction similar to the cold atom system is expected with lesser experimental complexity which could be advantageous for technological advancement for quantum information processing. 

\textbf{Methods} \\
The expression~(\ref{fit_f}) for $V_s$ was used to fit the non-linear phase shift of the probe measured in the experiment. To evaluate $V_s$, $\rho_{eff}$ is needed to be evaluated with
$N_b$ as the number of atoms in the blockade sphere. For the given vapor density and blockade radius, $N_b$ is not necessarily be an integer. Then, $N_{b}=N+x$ with $N$ being the integer 
and $x$ being the fractional number with $0\le x < 1$. $N+x$ can be written as $N\times \left(1-x\right)+\left(N+1\right)\times x$, which means
the probability of having $N$ atoms in the blockade sphere is $1-x$ and the probability of having $N+1$ atoms is $x$. The effective Rydberg population is modified as, 
$\rho_{eff}=\left(1-x\right)\times \rho_b\left(N\right)+x\times \rho_b \left(N+1\right)$ where $\rho_{b} (N)$ and $\rho_b(N+1)$ are evaluated using the expression~(\ref{emp1}). 
The above $\rho_{eff}$ is used to evaluate the integral in expression~(\ref{fit_f}) to find the fitting function of density. 

The non-linear phase shift measured using OHDT for different Rydberg states are shown in figure~\ref{fig3} and the respective errors are determined from repeated measurements. The vapor density was determined from the temperature of the vapor and respective errors were determined from the uncertainties in the measurement of the temperature. The dispersion data for $n=35$ at low density with $N_b<1$ were used to determine the overall gain $G$. Normal distribution of random numbers were generated for dispersion as well as density by taking their respective experimental errors as full width half maxima (FWHM). The statistical analysis of $G$ determined from the straight line fitting of the generated synthetic data gives mean value $\langle G\rangle =0.505$ and $1\sigma$ error $\delta{G} =0.01246$. Since all the experimental parameters including the probe and coupling Rabi frequencies were kept constant through out the experiment, then the same mean value and error for $G$ were used for further analysis of all the data. The same procedure was followed for all the data presented in figure~\ref{fig3} to fit with the expression~(\ref{fit_f}) by taking $r_b^3$ as the fitting parameter. Statistical analysis of the values of $r_b^3$ determined from the fitting of the synthetic data gives it's mean value and error as a function of principal quantum number which is shown in figure~\ref{fig4}. Again the the same procedure was followed to fit the data in figure~\ref{fig4} to determine the scaling law as $r^{3}_{b} \sim n^{* 5.8\pm 0.666}$.
 
\textbf{Acknowledgment}  We acknowledge Tanim Firdoshi, Snigdha S. Pati and Sushree S. Sahoo for assisting in performing the experiment. We are thankful to Dr. V. Ravi Chandra and 
 Dr. Anamitra Mukherjee for useful discussions regarding calculation for quantum many-body quantum system. This experiment was financially supported by the Department of 
 Atomic Energy, Govt. of India.

\textbf{Author contributions.} A.K.M. conceived the concept of blockade in thermal vapor. A.B. and D.K. performed the experiment.  A.B. analysed the experimental data. 
All the authors have contributed in theoretical modelling and preparation of the manuscript. A.B. and D.K. have contributed equally in this work.

\textbf{Data Availability Statement.} Correspondence and requests for data that supports the observation presented in this paper should be addressed to A.K.M. (email: a.mohapatra@niser.ac.in).

\textbf{Competing Interests:} The authors declare that they have no competing interests.
%=============================================================================================================
%=============================================================================================================
\begin{appendices}
\textbf{Interacting N-atom model}:\\
The Hamiltonian of N interacting atoms is given by $H=\sum_{i=1}^{N}\mathbf{I}^{i-1}\otimes H^{(i)}\otimes \mathbf{I}^{N-i}+\sum_{i<j}^{N}V_{ij}\ket{r}_{i}\ket{r}_{j}\bra{r}_{i}\bra{r}_{j}$ where $H^{(i)}$ represents the Hamiltonian of ith atom and $\mathbf{I}$ is the two dimensional identity matrix~\cite{prit10}. $V_{ij}$ represents the strength of the van der Waals interaction between atoms i and j. The decay and decoherence of the system can be included by the Lindblad operator given by $\mathbb{L}_{D}=\mathbb{L}_{D1}\otimes\rho^{(2)}\otimes \cdots \otimes \rho^{(N)}+\rho^{(1)}\otimes\mathbb{L}_{D2}\otimes\cdots\rho^{(N)}+\cdots+\rho^{(1)}\otimes\rho^{(2)}\otimes \cdots\otimes\mathbb{L}_{DN}$, where $\mathbb{L}_{Di}$ and $\rho^{(i)}$ represent the lindblad operator and the density matrix of ith atom~\cite{prit10}.
\section{2-atom model:}\label{2am}
\begin{figure}[h]
\includegraphics[scale=0.55]{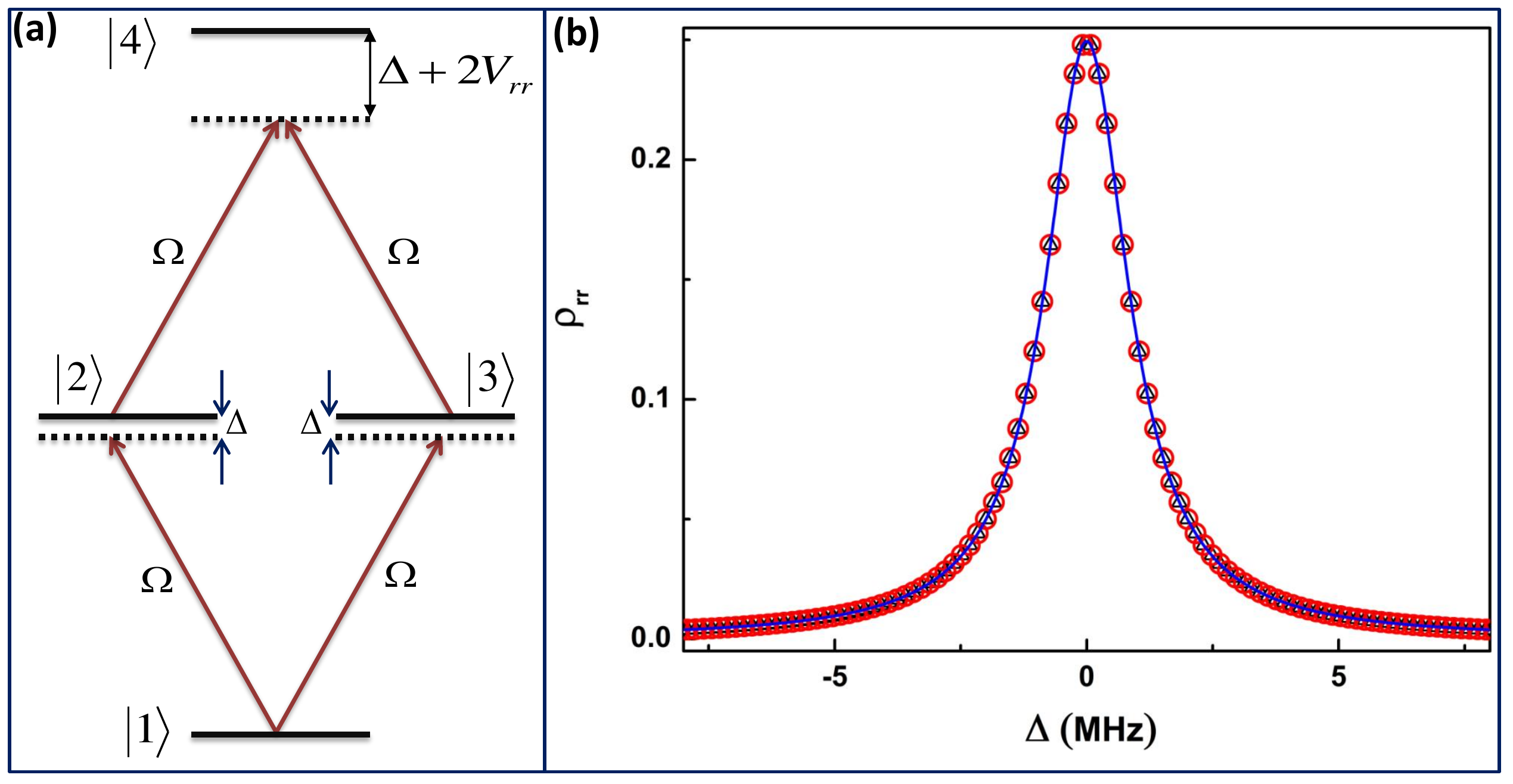}
\caption[]{(a) The energy level diagram of a two level system with levels $\ket{1}=\ket{gg}$, $\ket{2}=\ket{gr}$, $\ket{3}=\ket{rg}$ and $\ket{4}=\ket{rr}$. The applied laser is detuned by $\Delta$ from the atomic resonance and  the Rabi frequency is $\Omega$. (b) The population of the Rydberg state as a function of $\Delta$ with $\Omega=1$ MHz, $\Gamma=0.1$ MHz and $V_{rr}=10$ MHz for exact two atom calculation ($\triangle$), approximate model ($\circ$) and for the empirical formula (Solid line).}
\label{fig:2atom}
\end{figure}

Consider a system with two interacting identical atoms, each having two energy levels, a ground state $\ket{g}$ and an excited state $\ket{r}$ coupled by an applied laser with Rabi frequency $\Omega$. The energy level diagram of the composite system is depicted in figure~\ref{fig:2atom}(a). $\ket{1}$ corresponds to both the atoms in the ground state, $\ket{2}$ and $\ket{3}$ correspond to one atom in the ground state and the other in the Rydberg state and $\ket{4}$ corresponds to both the atoms in the Rydberg state. Thus, using the above expressions the Hamiltonian and the lindblad operator for the two atomic system can easily be constructed.  The master equation for the system as mentioned in the main text of the manuscript can be solved to calculate the Rydberg population $\rho_{rr}=\rho_{44}+(\rho_{22}+\rho_{33})/2$. The population of the Rydberg state as a function of the laser detuning for a strongly interacting 2-atom system satisfying the blockade condition is presented in Fig.~\ref{fig:2atom}(b). 

The system can further be simplified by considering the available symmetry and large energy level shift due to strong Rydberg-Rydberg interaction. The light  shift of the singly Rydberg excited state have been neglected. Coupling of the states $\ket{1}$ and $\ket{4}$ to the states $\ket{2}$ and $\ket{3}$ will be same. Hence, the population and the coherence terms will be the same, i.e. $\rho_{12}=\rho_{13}$, $\rho_{22}=\rho_{33}$ and $\rho_{24}=\rho_{34}$. Since the energy level shift of the state $\ket{4}$ is large due to Rydberg-Rydberg interaction, $\rho_{44}\approx0$ which leads to $\rho_{24}^{(im)}\approx0$. Using above approximations, the steady state optical Bloch equations (OBE) of the system can be simplified to a set of four equations as 

\begin{equation}
\begin{split}
\Omega(1-4\rho_{22}+\rho_{14})+2\Delta\rho_{12}+2i\Gamma(\rho_{24})-i\Gamma\rho_{12}&=0 \\
2\Omega(\rho_{12}-(\rho_{24}))+2(\Delta+V)\rho_{14}-2i\Gamma\rho_{14}&=0 \\
\Omega(Im(\rho_{12}))+\Gamma\rho_{22}&=0\\
\Omega(2\rho_{22}-\rho_{14})+2(\Delta+V)(\rho_{24})-3i\Gamma(\rho_{24})&=0
\end{split}
\end{equation}

These equations can be solved to evaluate the matrix element $\rho_{22}$ which gives the Rydberg population for the system. This Rydberg population is compared to that of the exact two atom model and is depicted in figure~\ref{fig:2atom}(b). The approximate model works well for a strongly interacting regime i.e. $V_{rr}\gg\sqrt{2}\Omega$. However, for $V_{rr}\thickapprox\sqrt{2}\Omega$ the approximate model deviates from the exact model. An empirical formula is derived using the super-atom model as discussed in the main text, for two atoms inside the blockade sphere and is given by, $\rho_{rr}=\dfrac{\Omega^2}{4\Omega^2+4\Delta^2+\Gamma^2}$. Rydberg population calculated using super atom model is also depicted in figure~\ref{fig:2atom}(b). The empirical formula matches well with the exact model and the approximate model.

\section{Three atom system}\label{3am}  
\begin{figure}
\includegraphics[scale=0.45]{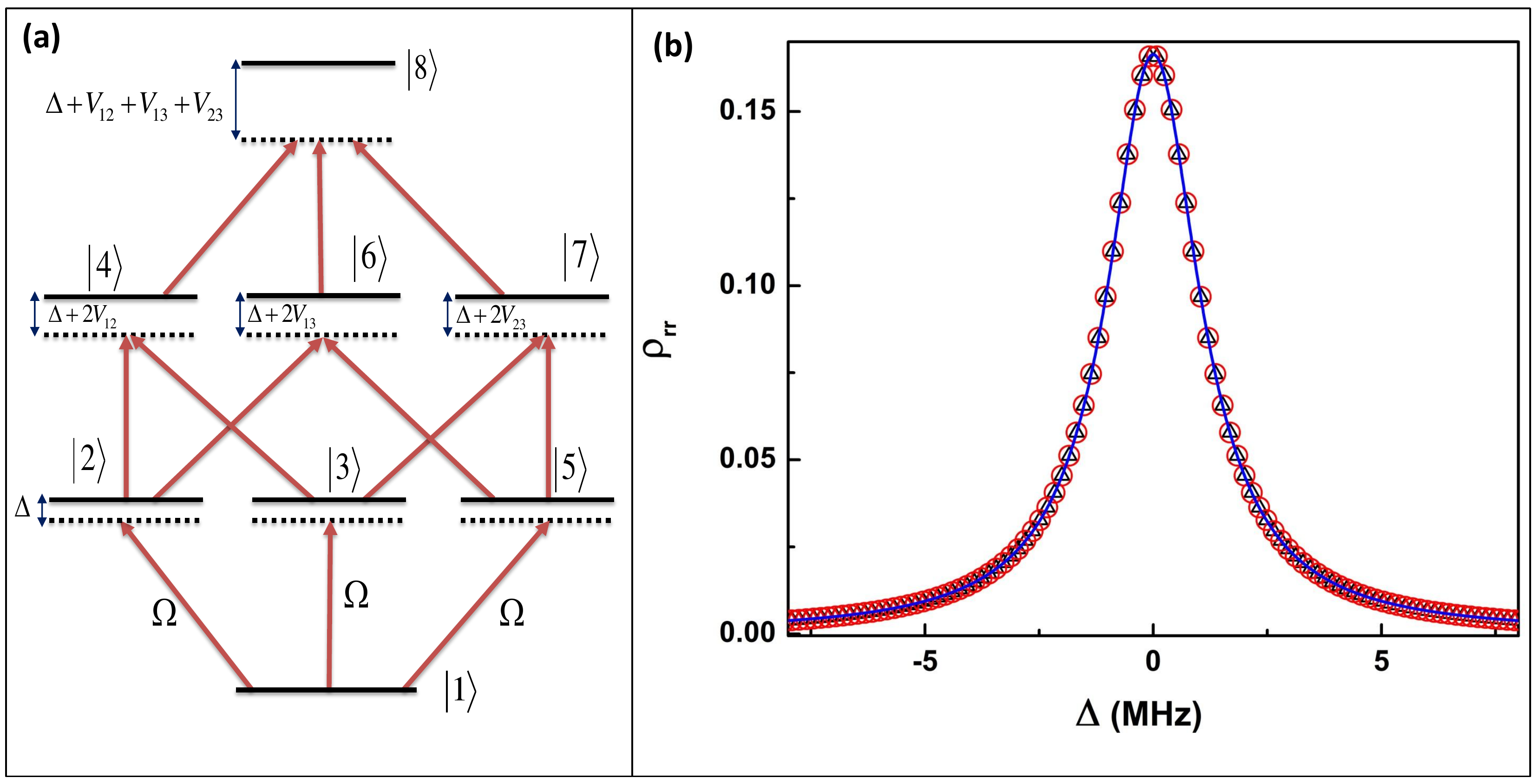}
\caption[]{(a) The energy level diagram for three interacting atoms. The applied laser is detuned by $\Delta$ from the atomic resonance and  Rabi frequency of the transition is $\Omega$. (b) The population of the Rydberg state as a function of $\Delta$ with $\Omega=1$ MHz, $\Gamma=0.1$ MHz and $V_{rr}=10$ MHz for exact three atom calculation ($\triangle$), approximate model ($\circ$) and for empirical formula (solid line).}
\label{fig:3atom}
\end{figure}

Now consider a system with three interacting identical atoms, each having two energy levels, a ground state $\ket{g}$ and an excited state $\ket{r}$ coupled by an applied laser with Rabi frequency $\Omega$. The energy level diagram of the composite system is depicted in fig.~\ref{fig:3atom}(a). $\ket{1}$ corresponds to all the atoms in the ground state. $\ket{2}$, $\ket{3}$ and $\ket{5}$ represent any one atom in Rydberg state while other two in the ground state. $\ket{4}$, $\ket{6}$ and $\ket{7}$ represent states with two atoms in the Rydberg state with one atom in the ground state and $\ket{8}$ represents all the atoms in the Rydberg state. The Hamiltonian and the lindblad operator for the three atom system can be calculated using the expression for N atom presented in the previous section. The OBE of the system can be solved numerically in steady state to calculate the Rydberg population as $\rho_{rr}=\dfrac{1}{3}(\rho_{22}+\rho_{33}+\rho_{55})+\dfrac{2}{3}(\rho_{44}+\rho_{66}+\rho_{77})+\rho_{88}$. The population of the Rydberg state as a function of laser detuning is depicted in fig.~\ref{fig:3atom}(b).

The model can further be simplified using the available symmetry in the system and large energy level shift due to strong Rydberg interaction. We have ignored the inhomogeneous light shift of the single Rydberg excited state which behave like dephasing in many-body Rabi oscillation~\cite{dudi12}. Coupling of the  states $\ket{1}$, $\ket{4}$, $\ket{6}$, $\ket{7}$ and $\ket{8}$ to the states $\ket{2}$, $\ket{3}$ and $\ket{5}$ will be the same. Hence the population and the corresponding coherence term will be same, i.e. $\rho_{12}=\rho_{13}=\rho_{15}$, $\rho_{22}=\rho_{33}=\rho_{55}$, $\rho_{24}=\rho_{26}=\rho_{34}=\rho_{37}=\rho_{56}=\rho_{57}$, $\rho_{28}=\rho_{38}=\rho_{58}$ and $\rho_{27}=\rho_{36}=\rho_{54}$ and $\rho_{16}=\rho_{14}=\rho_{17}$. The shift of the states $\ket{4}$, $\ket{6}$, $\ket{7}$ and $\ket{8}$ will be large due to strong interaction between the atoms. Hence the populations $\rho_{44}$=$\rho_{66}$=$\rho_{77}$=$\rho_{88}$=0. Thus coherence between these states will be zero, i.e. $\rho_{48}=\rho_{68}=\rho_{78}=\rho_{46}=\rho_{47}=\rho_{67}=0$. Thus implementing all these conditions, the three atom system in steady state can be reduced to a set of 4 equations as 
\begin{equation}
\begin{split}
 \Omega(1-6\rho_{22}+2\rho_{14})+2\Delta\rho_{12}+4i\Gamma\rho_{24}-i\Gamma\rho_{12}=0\\
 2\Omega(\rho_{12})+2(2\Delta+V)\rho_{14}-2i\Gamma\rho_{14}=0\\
 \Omega(Im\rho_{12})+\Gamma\rho_{22}=0\\
 \Omega(2\rho_{22}-\rho_{14})+2(\Delta+V)(\rho_{24})-3i\Gamma\rho_{24}=0\\
 \end{split}
\end{equation}  
These equation can be solved to calculate the matrix element $\rho_{22}$ which gives the Rydberg population. This population can be compared to the Rydberg population of the exact three level system and is depicted in Fig.~\ref{fig:3atom}(b). The approximate model is found to match well with the exact model for a strongly interacting regime as mentioned in the two atom model. However with an interaction strength $V_{rr}\thickapprox\sqrt{3}\Omega$ the appoximate model deviates from the exact calculation. The empirical formula derived from the super atom model as mentioed in the main text having three atoms inside the blockade sphere is given by $\rho_{rr}=\dfrac{\Omega^{2}}{6\Omega^{2}+4\Delta^{2}+\Gamma^{2}}$. Rydberg population calculated from the super atom model is also depicted in Fig.~\ref{fig:3atom}(b). The empirical formula matches well with the approximate calculation and also with the exact three atom model with $V_{rr} \gg \sqrt{3}\Omega$.
 
\section{N-atoms}\label{nam}
 \begin{figure}[h]
\includegraphics[scale=0.4]{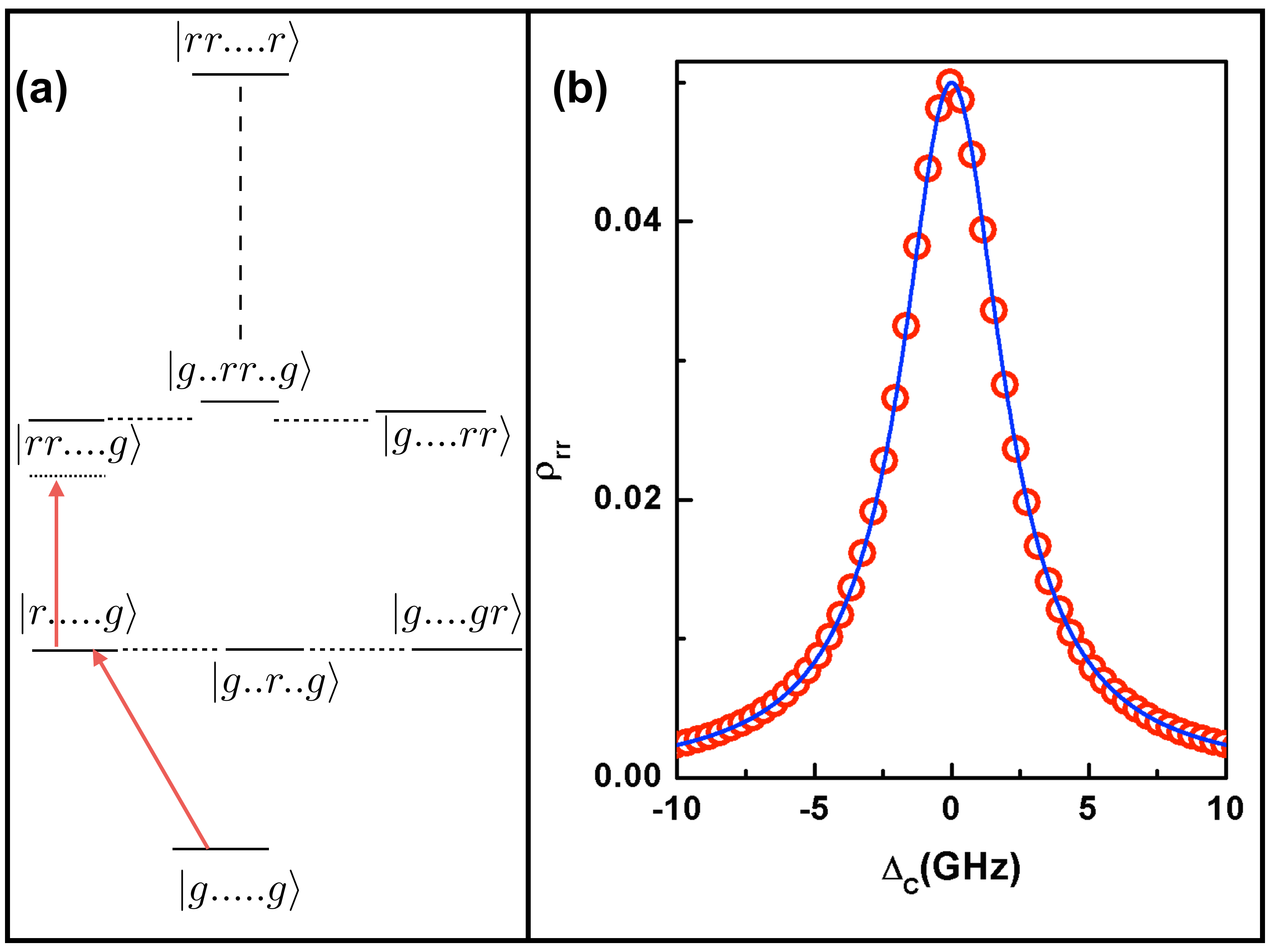}
\caption{(a) The energy level diagram for N interacting atoms with states $\ket{1}=\ket{ggg\cdot\cdot\cdot\cdot g}$, $\ket{2}=\ket{rgg\cdot\cdot\cdot\cdot g}$ and $\ket{3}=\ket{rrg\cdot\cdot\cdot\cdot g}$. (b) The population of the Rydberg state with laser detuning $\Delta$ with $\Omega=1$ MHz, $\Gamma=0.1$ MHz with $V_{rr}=20$ MHz for approximate model ($\circ$) and for the empirical formula (line).}
\label{fig:10atom}
\end{figure}
Consider a system of  N interacting identical atoms with energy level diagram as depicted in Fig.~\ref{fig:10atom}(a). The hamiltonian and the lindblad operator can be written using the expressions presented in the first section. From the 2-atom and 3-atom calculation presented above, we observe that both of them can be reduced to an effective model with set of four independent OBE. These two sets of equations can be extrapolated for N atoms and is given by,
\begin{center}
$\Omega(1-2N\rho_{22}+(N-1)\rho_{13})+2\Delta\rho_{12}+2i\Gamma\rho_{23}-i\Gamma\rho_{12}=0$\\
$2\Omega(\rho_{12})+\Omega(N-3)\rho_{23})+2((N-1)\Delta+V)\rho_{13}-2i\Gamma\rho_{13}=0$\\
$\Omega(Im(\rho_{12}))+\Gamma\rho_{22}=0$\\
$\Omega(2\rho_{22}-\rho_{13})+2(\Delta+V)(\rho_{24})-3i\Gamma\rho_{23}=0$
\end{center} 
The population of the Rydberg state is given by $\rho_{22}$. The population as a function of laser detuning is depicted in Fig.~\ref{fig:10atom}(b) for N=10. The empirical formula of Rydberg population from the super atom model with N atoms in the blockade sphere, as mentioned in the main text is given by $\rho_{rr}=\dfrac{\Omega^{2}}{2N\Omega^{2}+4\Delta^{2}+\Gamma^{2}}$. A comparison of the Rydberg population calculated from  the empirical formula and the approximate model for 10 atoms is depicted figure~\ref{fig:10atom}(b). A good agreement is observed between both the models.  

Thus a system of N interacting atoms in the blockade sphere can be reduced to a set of four equation using the available symmetry and strong Rydberg Rydberg interaction. We observe that for  $V_{rr}\gg\sqrt{N}\Omega$ a nice match is observed between the superatom model and the approximate model. However for $V_{rr}\thickapprox\sqrt{N}\Omega$ the super-atom model deviates from approximate calculation.
\end{appendices}
%=============================================================================================================
%=============================================================================================================

\end{document}